\documentclass[twocolumn]{aastex63}

 \hypersetup{linkcolor=cyan,citecolor=cyan,filecolor=cyan,urlcolor=blue}

\usepackage{graphicx}
\usepackage{amsmath}
\accepted{for publication in the Astrophysical Journal, September 10, 2020}

\shorttitle{Quasi-adiabatic and stochastic heating}
\shortauthors{K. Stasiewicz \& B. Eliasson}

\begin{document}

\title{Quasi-adiabatic and stochastic heating and particle acceleration at quasi-perpendicular shocks}


\author[0000-0002-2872-5279]{Krzysztof Stasiewicz}
\email{krzy.stasiewicz@gmail.com}
\affiliation{Department of Physics and Astronomy, University of Zielona G\'ora, Poland }
\affiliation{Space Research Centre, Polish Academy of Sciences, Warsaw, Poland }
\author[0000-0001-6039-1574]{Bengt Eliasson}
\email{bengt.eliasson@strath.ac.uk}
\affiliation{SUPA, Department of Physics, University of Strathclyde, Glasgow, G4 0NG, United Kingdom}

\begin{abstract}
Based on Magnetospheric Multiscale (MMS) observations from the Earth's bow shock, we have identified two plasma heating processes that operate at quasi-perpendicular shocks. Ions are subject to stochastic heating in a process controlled by the heating function  $\chi_j = m_j q_j^{-1} B^{-2}\mathrm{div}(\mathbf{E}_\perp)$ for particles with mass $m_j$ and charge $q_j$ in the electric and magnetic fields $\mathbf{E}$ and $\mathbf{B}$. Test particle simulations are employed to identify the parameter ranges for bulk heating and stochastic acceleration of particles in the tail of the distribution function. The simulation results are used to show that ion heating and acceleration in the studied bow shock crossings is accomplished by  waves  at frequencies (1--10)$f_{cp}$ (proton gyrofrequency) for the bulk heating, and $f>10f_{cp}$  for the tail acceleration. When electrons are not in the stochastic heating regime, $|\chi_e|<1$, they undergo a quasi-adiabatic heating process characterized by the isotropic temperature relation $T/B=(T_0/B_0)(B_0/B)^{1/3}$. This is obtained when the energy gain from the conservation of the magnetic moment is redistributed to the parallel energy component through the scattering by waves. The results reported in this paper may also be applicable  to particle heating and acceleration at astrophysical shocks.

\end{abstract}



\section{Introduction} \label{seq1}
Electron and ion acceleration and heating at collisionless shocks is an important problem in astrophysics and space physics, which has been addressed over many years by a number of authors
\citep{Bell:1978,Lee:1982,Wu:1984,Goodrich:1984,Blandford1987,Balikhin:1994,Gedalin:1995,Treumann:2009,Burgess:2012,See:2013,Mozer:2013,Krasnoselskikh:2013,Guo2014,Wilson:2014,Park2015,Cohen:2019,Xu2020}. In the above cited publications one can find a long list of  waves, instabilities, and processes that could  play a role in ion and electron heating/acceleration -- however, the question of the exact heating mechanisms working at shocks is still in an inconclusive state.

New generation of  high-resolution space instruments with simultaneous 4-spacecraft  measurements on the Magnetospheric Multiscale (MMS) mission \citep{Burch:2016} opened unprecedented  possibility for testing heating and acceleration mechanisms that  operate at collisionless shocks in reality, and not only in theory. For example, MMS offers the capability of computing gradients of plasma parameters and of electric and magnetic fields on spacecraft separation distances of $\sim 20$ km, equivalent to several electron gyroradii. The quality of the electric field experiment \citep{Lindqvist:2016,Ergun:2016,Torbert:2016} enables  even the direct derivation of the divergence of the electric field.  Using such state-of-the-art measurements, which will also be discussed in Section \ref{sec2} of the present paper, \citet{Stasiewicz:2020a,Stasiewicz:2020b} has identified a chain of cross-field current driven instabilities that  operate at both quasi-parallel and quasi-perpendicular shock waves and lead to the heating of ions and electrons. This sequence can be summarized as follows:

{\em  Shock compression of  the density N and  the magnetic field B }    $\rightarrow$ {\em  diamagnetic current}  $\rightarrow$ {\em lower hybrid drift (LHD) instability} $\rightarrow$  {\em electron ${\bf E}\times {\bf B}$ drift} $\rightarrow$ {\em electron cyclotron drift (ECD) instability}   $\rightarrow$ {\em heating: quasi-adiabatic ($\chi_j<1$), stochastic  ($\chi_j>1$)}.

Stochastic heating is a single particle mechanism where large electric field ${\bf E}$ gradients due to space charges destabilize individual particle motions in a magnetic field ${\bf B}$, rendering the trajectories chaotic in the sense of a positive Lyapunov exponent for  initially nearby states. The  stochastic heating function of particle species $j$ ($j=e$ for electrons and $p$ for protons) is \citep{Stasiewicz:2020a}

\begin{equation}
\chi_j(t,\mathbf{r})  = \frac{ m_j}{q_j B^2} {\rm div}(\mathbf{E}_\perp)  \label{eq1}
\end{equation}
where $m_j$ and $q_j$ are the particle mass and charge. The parallel (to the magnetic field) electric field ${\bf E}_\parallel$ is here excluded since it does not directly contribute to the stochasticity, leaving only the perpendicular field ${\bf E}_\perp$ in (\ref{eq1}).
Stochastic heating typically occurs when $|\chi_j |\gtrsim1$ \citep{Karney:1979,McChesney:1987,Balikhin:1993,Gedalin:1995,Vranjes:2010,Stasiewicz:2013,See:2013,Yoon:2019}  even though resonant heating can also occur for $|\chi_j |<1$ for wave frequencies very close to cyclotron harmonics \citep{Fukuyama:1977}. The value of $\chi_j$ can be regarded as a measure of demagnetization. Particles are magnetized (adiabatic) for $|\chi_j|<1$, and demagnetized (subject to non-adiabatic heating) for $|\chi_j|\gtrsim1$.

The value of the proton heating function $\chi_p$ is typically in the range $10-100$ in the bow shock and the magnetosheath, which implies that the ions are strongly demagnetized and can be subject to stochastic heating processes in these regions. In Section \ref{sec4}, test-particle simulations are carried out for a range of parameters primarily relevant for stochastic heating of protons.

In order to also demagnetize and stochastically heat electrons we need $\chi_e>1$ and therefore $\chi_p>m_p/m_e=1836$, which  requires either very strong $E$-gradients or low $B$-fields, or both, as implied by Eq.~(\ref{eq1}).
Electron heating at perpendicular shocks based on $\chi_e$ (with the divergence reduced to $\partial E_x/\partial x$) has been referred to as the \emph{kinematic mechanism} \citep{Balikhin:1993,Gedalin:1995,See:2013}. The required gradient of the electric field is associated with a macroscopic electric field in the direction normal to the shock. Unfortunately, in perpendicular shocks the observed thickness of the shock ramp and measured values of the normal electric field do not allow $\chi_e>1$ to be reached, as needed for stochastic heating of electrons with the kinematic scenario. On the other hand, the stochastic heating mechanism has been shown to work with gradients of the electric field provided by the LHD and ECD waves observed in quasi-parallel shocks \citep{Stasiewicz:2020b}.

In quasi-perpendicular shocks the  derived values of $\chi_e$ are mostly below the stochastic threshold for electrons, because of the increasing values of $B\approx10-40$ nT in the shock ramp, combined with the scaling $\chi_e \propto B^{-2}$. In Section \ref{sec3}, we demonstrate that such situations instead lead to \emph{quasi-adiabatic electron heating}, characterized by electron heating on the compression of the magnetic field, combined with  scattering by waves, leading to the isotropic temperature relation $T/B=(T_0/B_0)(B_0/B)^{1/3}$. This is to our knowledge  a novel concept identified and explained for the first time in Section \ref{sec3}.

\section{Multiple  crossings and waves in shocks} \label{sec2}

We analyse recent MMS measurements from 2020-01-03 obtained by the 3-axis  electric field \citep{Lindqvist:2016,Ergun:2016,Torbert:2016} and  magnetic field vectors measured by the Fluxgate Magnetometer \citep{Russell:2016},  and the number density, velocity, and temperature of both ions and electrons from the Fast Plasma Investigation  \citep{Pollock:2016}.

\begin{figure}
\includegraphics[width=\columnwidth]{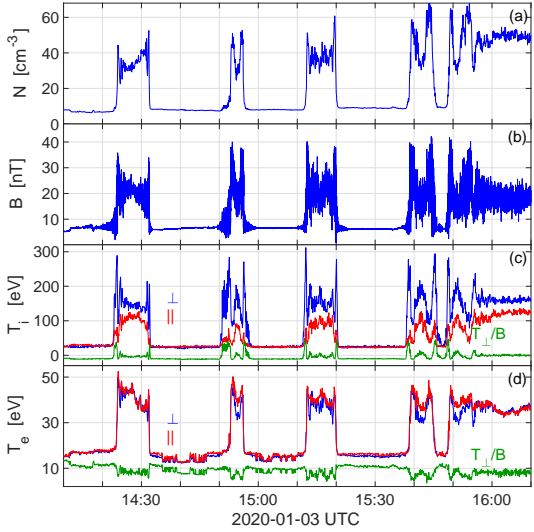}
\caption{A series of  shock crossings caused by the oscillatory movement of the bow shock. Panel (a) shows the electron number density $N$, panel (b) the magnetic field $B$, panel (c) shows  $T_{i\perp}, T_{i\parallel}$, and the ratio $T_{i\perp}/B$ (not to scale), and panel (d) shows the same parameters as (c) but for electrons. Note the different behaviors of the ratio $T_\perp/B$ for the ions and electrons in the shock ramps, with humps for the ions and dips for the electrons. \label{Fig1}}
\end{figure}

Figure~\ref{Fig1} shows  multiple crossings of   shocks caused by the oscillatory movements of the bow shock with an amplitude of 6--10 km/s estimated from the time shifts of the density signals. The speed is with respect to the MMS spacecraft moving at 1.9 km/s earthward. The spacecraft position at time 14:30 was (13.5, 10.3, -1.8) R$_E$ GSE (geocentric solar ecliptic) coordinates, and the average inter-spacecraft distance was 22 km (minimum 10 km). Shown in Fig.~\ref{Fig1} are: the electron number density $N$, the magnetic field $B$, the ion and electron temperatures, and the ratio $T_{\perp}/B$ (not to scale).     Notably in Fig.~\ref{Fig1}c, the parallel and perpendicular electron temperatures are almost equal, indicating that an isotropization process takes place.

\begin{figure}
\includegraphics[width=\columnwidth]{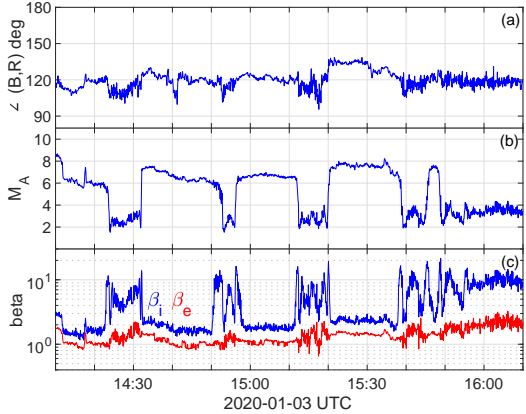}
\caption{Complementary plasma parameters for Fig.~\ref{Fig1}: (a) the angle between the magnetic field vector and the geocentric radial direction (a proxy to the shock normal direction), (b) the  Alfv\'{e}n Mach number, and (c) ion and electron $\beta$ parameter. \label{Fig2}}
\end{figure}

Complementary plasma parameters for the shock crossings are displayed in Fig.~\ref{Fig2}, which shows (a) the  angle between the magnetic field vector and the geocentric  radial direction to the spacecraft (a proxy for the shock normal), (b) the Alfv\'{e}n Mach number, $M_A=V_i/V_A$, i.e, the ratio between the ion bulk speed and the Alfv\'{e}n speed, and (c) the ion and electron plasma beta, i.e., the ratio between the thermal energy density of particles and the magnetic field energy density.  It is seen that all shocks have  quasi-perpendicular configurations, the plasma beta is $\beta_i\sim 2, \; \beta_e\sim 1$, and the Alfv\'{e}n Mach number 6-8, outside the shocks.

The ion and electron temperatures indicate rapid heating at the shock ramp, and the repetitive events offer a great opportunity to  study   heating processes operating at quasi-perpendicular shocks. The  ratio $T_{\perp}/B$  derived from measurements is an excellent indicator of the heating processes. A flat ratio across the shock would indicate adiabatic perpendicular heating coming from the conservation of the magnetic moment. This should be accompanied by unchanged parallel temperature.
We see that the ion ratio $T_{i\perp}/B$ has humps and the parallel temperature is smaller, which is indicative for non-adiabatic perpendicular heating and less efficient parallel heating. The electron ratio $T_{e\perp}/B$ instead has dips and the temperature is nearly isotropic.

\begin{figure}
\includegraphics[width=\columnwidth]{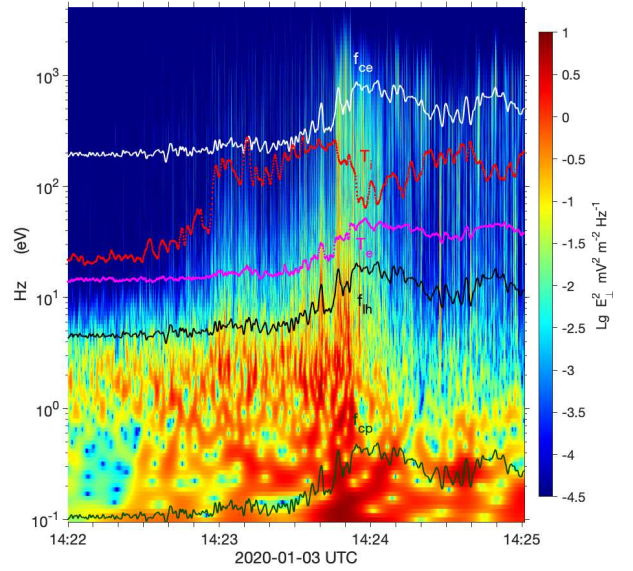}
\caption{Time-frequency spectrogram of the perpendicular electric field for the first shock ramp in Fig. \ref{Fig1}.
Over-plotted are the proton cyclotron frequency $f_{cp}$, the lower hybrid frequency $f_{lh}$, the electron  and ion temperatures (eV), and the electron cyclotron frequency $f_{ce}$. Waves between $f_{cp}-f_{lh}$  are attributed to the LHD and above $f_{lh}$ to the MTS (modified two-stream) instabilities, and for frequencies around $f_{ce}$ and above to the ECD instability. Note the vertical striations that start from below 1 Hz (LHD instability) and go through the MTS and ECD instabilities up to 4 kHz, indicating co-location and common origin of these instabilities. \label{Fig3}}
\end{figure}

 Both particle species  in Figure~\ref{Fig1} manifest non-adiabatic behavior but of a different character.
 To study the processes and wave modes involved in the heating we show in Fig.~\ref{Fig3} the time-frequency power spectrogram of the perpendicular electric field sampled at the rate 8192 s$^{-1}$  for the first shock crossing.
Over-plotted are the  proton cyclotron frequency $f_{cp}$, the lower hybrid frequency $f_{lh}$, the electron temperature $T_{e\perp}$, the ion temperature $T_{i\perp}$, and the electron cyclotron frequency $f_{ce}$.  Near the peak of the shock at around 14:24 UT, the mean values of the proton plasma frequency is $f_{pp}\approx 1$ kHz,  and the electron plasma frequency $f_{pe}\approx 42$ kHz.
The lower-hybrid frequency is $f_{lh}=[f_{pp}^{-2}+(f_{cp}f_{ce})^{-1}]^{-1/2}\approx 10$ Hz.

Waves  below   $f_{lh}$ are  related to the LHD instability, which has maximum growth rate at $k_\perp r_e\sim 1$  \citep{Davidson:1977,Drake:1983}, however simulation results of \citet{Daughton:2003} indicate that they have longer wavelengths $k_\perp(r_e r_p)^{1/2}\sim 1$ and electromagnetic character with significant magnetic fluctuations, which are observed also in the present case. Here, $k_\perp$ is the wavenumber perpendicular to the magnetic field, and $r_e=v_{Te}/\omega_{ce}$ is the electron thermal Larmor radius, and $v_{Te}=(T_{e\perp}/m_e)^{1/2}$ is the electron thermal speed, $r_p$ is the proton thermal Larmor radius. For $r_e\approx 1$ km, and ion perpendicular speed $V_{i\perp}\approx 250$ km/s with respect to the spacecraft, the LHD waves at frequencies $f<f_{lh}$ and wavelengths $\lambda_M\sim 2\pi r_e$ corresponding to the maximum growth rate  will be upshifted in frequency by $\Delta f= V_{i\perp}/\lambda_M$ and observed up to 50 Hz in Figure \ref{Fig3}. Notably, there are other wave modes in this frequency range, as for example magnetosonic whistlers at $\sim$2 Hz that propagate upstream from the shock at 14:22 UTC, and are seen also in Fig.~\ref{Fig1}b.  Waves around $f_{ce}$ and above are associated with the ECD instability and coupled with ion-acoustic (IA) waves \citep{Wilson:2010,Breneman:2013,Goodrich:2018,Stasiewicz:2020a}. Because of short wavelengths $\sim  r_e$ they could be Doppler  downshifted by up to hundreds Hz.

The LHD and ECD instabilities are cross-field current driven instabilities caused by relative electron-ion drifts. Diamagnetic drift of protons   $V_{d}=T_p(m_p\omega_{cp}L_N)^{-1}=v_{Tp}(r_p/L_N)$ caused by gradients of the density  leads to the LHD instability when the ratio between the scale of the density gradient $L_N=N|\nabla N|^{-1}$ and  the proton thermal gyroradius $r_p$ obeys the condition $L_N/r_p<(m_p/m_e)^{1/4}$  \citep{Huba:1978,Drake:1983}.
 The density compression starts at the foot of the shock and is strongly amplified in the ramp, which can be seen in Figure \ref{Fig3} in the profile $f_{lh}$, which is proportional to $B$, but also representative for $N$.

In the nonlinear stage, the LHD waves produce large amplitude electric fields resulting in efficient ${\bf E} \times {\bf B}$ drifts of the electrons. Due to the large ion gyroradius compared to the wavelength of the LHD waves, the ions do not experience significant ${\bf E} \times {\bf B}$, and hence there is a net current set up by the electrons only. When the differential drift velocity between the electrons and ions exceeds the ion thermal velocity, the modified two-stream  (MTS) instability  can take place resulting in waves at frequencies above $f_{lh}$ \citep{Krall:1971,Lashmore:1973,Gary:1987,Umeda:2014}. Below, we will not differentiate between the MTS and LHD instabilities since they belong to the same dispersion surface \citep{Silveira:2002,Yoon:2004}, but we will use the term LHD instability in the sense of a generalized cross-field current driven instability in the lower-hybrid frequency range.

\begin{figure}
\includegraphics[width=\columnwidth]{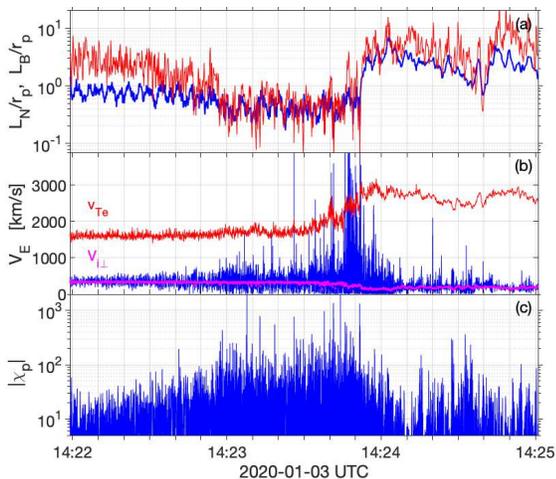}
\caption{Diagnostic parameters for the case in Fig.~\ref{Fig3}: (a) The gradient scales of the plasma density $L_N$  (blue) and of the magnetic field $L_B$ (red) are normalized with thermal proton gyroradius $r_p$ (mean = 91 km). Regions with  $L_N/r_p \lesssim 1$ are  unstable for the LHD instability.  (b) The computed ${\bf E}\times {\bf B}$ drift speed $V_E$ (blue), the electron thermal speed $v_{Te}$ (red), and perpendicular ion speed $V_{i\perp}$ (magenta). Regions with $V_E \sim v_{Te}$  indicate presence of the ECD instability.   (c) The stochastic heating function $\chi_p$  for protons derived from the data with Eq. (\ref{eq1}) for the electric field 0.25--512 Hz.  \label{Fig4}}
\end{figure}

When the relative electron-ion drift speed becomes a significant fraction of the electron thermal speed, $V_{E}=|{\bf E}\times{\bf B}|/B^2\sim v_{Te}$, the ECD instability is initiated, which creates even larger electric fields on spatial scales of $r_e$. \citep{Forslund:1972,Lashmore:1971,Muschietti:2013}.
The  ECD instability takes place near cyclotron resonances $(\omega - k_\perp V_{de})= n\omega_{ce}$, where $\omega_{ce}=eB/m_e$ is the angular electron cyclotron frequency, $k_\perp=2\pi/\lambda$ is the perpendicular wave number, $\lambda$ is the wavelength, and $n$ is an integer \citep{Janhunen:2018}. Here, $V_{de}\approx V_E$ is the perpendicular electron drift velocity in the rest frame of the ions. This resonance condition can be written $k_\perp V_{E} \approx n\omega_{ce}$ and expressed as $k_\perp r_e \approx nv_{Te}/V_E$. For $r_e=1$ km and $n=1$ their wavelengths are

 \begin{equation}
 \lambda \approx \frac{2\pi r_e}{n} \frac{V_E}{v_{Te}} \approx 6.3\,[\mathrm{km}]\, \frac{V_E}{v_{Te}}. \label{lambda}
 \end{equation}
This  means that the ECD waves with $n=1$ and electric drift velocities $V_E>v_{Te}$  (see Figure \ref{Fig4}b) have wavelengths large enough to enable correct gradient computations in the calculations of div($\mathbf{E}$) by the MMS spacecraft constellation. Contribution of shorter waves with $n>1$ to the computed $\chi$ may be underestimated by the gradient computation procedure.
The ECD waves resonate/couple with structures created by the LHD instability, $k_\perp r_e\sim 1$, when $nv_{Te}/V_E=1$. The $n$=1  ECD mode can be naturally excited in drift channels created by the LHD instability when $V_E=v_{Te}$. There is smooth transition and co-location of LHD and ECD waves, seen in Fig. \ref{Fig3} which is possibly related to the matching condition  between these two instabilities.

The  scale of the density gradient $L_N$   shown in Fig. \ref{Fig4}a  is computed directly from 4-point measurements using the method of  \citet{Harvey:1998}.
 As a verification, we show also the gradient scale $L_B=B|\nabla B|^{-1}$ for the magnetic field. They coincide in the shock proper, as expected  for fast magnetosonic structures.

In the pioneering work on the LHD instability, \citet{Krall:1971} used an expression for the electron drift  in the form $V_{de} \propto (N^{-1}\partial_x N - B^{-1} \partial_x B + ....  )$, which implied that the current due to the $\nabla B$ term from fluid integration would cancel the diamagnetic current due to the $\nabla N$ drift, in case when the gradient scale lengths are the same ($L_N=L_B$) and both gradients point in the same direction.
Because these scales are the same at the bow shock, several influential authors \citep{Lemons:1978,Zhou:1983,Wu:1984}, and many others, claimed that there would be no LHD instability at the bow shock. This erroneous conclusion affected many researchers afterwards and has led to a 40-year delay in the identification of the LHD instability as the prevailing ion heating mechanism in  compressional shock waves, and as a possible trigger for the ECD instability \citep{Stasiewicz:2020a,Stasiewicz:2020b}.  As a matter of fact, in a  homogeneous plasma with a spatially varying magnetic field,  the $\nabla B$ drift cancels with other terms due to the gyration of particles in the magnetic field, and therefore does not contribute to macroscopic currents as explained in section 7.4 of the textbook by \citet{Goldston:1995}.  Thus, the diamagnetic drift current is not canceled by the magnetic gradient drift term, and the LHD instability can be excited at the bow shock.

Figure \ref{Fig4}b shows the computed ${\bf E} \times {\bf B}$ drift speed which is increased and  comparable to the electron thermal speed in the ion and electron heating regions in Fig. \ref{Fig3}. The drift velocity was computed in the frequency range 0 - 512 Hz, because for frequencies larger than the electron cyclotron frequency the drift approximation is not valid.  For comparison we also show  the plot of the measured perpendicular ion speed $V_{i\perp}$ in magenta color. It has the same value as the computed ${\bf E}\times {\bf B}$ drift speed in the solar wind up to 14:23, but deviates strongly inside the shock.  Large difference between the electron drift $V_E$ and the measured perpendicular drift of ions $V_{i\perp}$ would induce sequentially the LHD and ECD instabilities as mentioned in the Introduction.
The diagnostic parameters support the interpretation of the waves shown in Fig. \ref{Fig3} as caused by the LHD and ECD instabilities, and that these waves are spatially co-located, which is seen as striations extending over the whole frequency spectrum.

Stochastic heating is controlled by the stochastic heating function (\ref{eq1}) computed directly from 4-point measurements using the method of  \citet{Harvey:1998}.
 \citet{Goodrich:2018} raised concerns that the axial double probe (ADP) on MMS, which uses rigid axial booms shorter than the wire booms of the spin-plane double probe experiment (SDP), produces a larger amplitude response for short, tens of meter (Debye length) waves such as the ion-acoustic (IA) wave. This instrumental difference may affect the computations of the divergence of the electric field and the resulting value of $\chi$. Therefore, the computations of div(\textbf{E}) are made in the despun spacecraft coordinates (DSL), which separates $E_z$ provided by the ADP, from ($E_x,E_y$) provided by the SDP. This enables  removal of the highest frequency components above $f_{ce}$, which may contain such short waves around the ion plasma frequency $f_{pp}$, from the analysis. We have therefore removed the highest frequency components before computing the gradients, and $\chi_p$ shown in Fig.~\ref{Fig4}c is computed for the frequency range 0.25--512 Hz.   Frequencies below 0.25 Hz are removed to avoid spurious effects at the spin frequency and its harmonics. Not using the ADP $E_z$ component at all produces $\chi$  ca 20\% smaller.

The computed $\chi_p$   shown in Fig. \ref{Fig4}c  indicates that the ions are demagnetized and likely to undergo stochastic heating as seen in detail in Fig. \ref{Fig3}. On the other hand the value of $\chi_p\sim 100$ corresponds to $\chi_e\sim 0.06$, which is too small to demagnetize the electrons and subject them to stochastic heating.
 The computed contribution to $\chi_e$ from  short $n>1$ ECD waves (Eq. \ref{lambda}) is underestimated  -- however, even a possible correction  would still keep it below the stochasticity threshold. Other errors in the derivation of $\chi$ are the same as in measurements of the electric field, i.e., 1 mV/m, or ca 10\% for large amplitude fields \citep{Lindqvist:2016}.

Figure \ref{Fig3} is representative for all 9 shock crossings shown in Fig. \ref{Fig1}. It shows that the  ion heating is observed at the foot of the shock, earlier than the electron heating, and  correlates well with the power of LHD waves. Electron temperature correlates well with the increased wave activity in the LH/ECD frequency range and maximizes in the region of the most intense ECD waves around  14:24.
On the other hand, it appears to correlate also with the magnetic field strength, represented here by the electron gyrofrequency $f_{ce}=\omega_{ce}/2\pi$, which suggests that an adiabatic behavior $T_{e\perp}\propto B$ should be also considered here. However, this apparent correlation is not exact, as seen in the $T_{e\perp}/ B$ ratio shown in Fig. \ref{Fig1}, with humps for the ions and dips for the electrons at the shock ramps. This will be explained in the next section as quasi-adiabatic electron heating involving the compression of the magnetic field combined with isotropization by the scattering on waves.

\section{Quasi-adiabatic electron heating} \label{sec3}

The computed value of the heating function (\ref{eq1}) shown in Fig. \ref{Fig4}c indicates that the stochastic heating may not be available for electrons in the analyzed shock crossings.
The behavior of the ratio $T_{e\perp}/B$ and the isotropy of the electron temperature, discussed in Section \ref{sec2}, suggests a different kind of heating process.  Let us assume that the electrons obtain perpendicular energy from the conservation of the magnetic moment (they are magnetized, consistent with $\chi_e<1$), but  the energy gain is redistributed to the parallel component through the scattering by waves.

When the magnetic moment is conserved, i.e., $T_\perp/B=const$, the differential temperature increase is $dT_\perp=T_\perp B^{-1}dB$. If the energy gain from 2 degrees of freedom ($2dT_\perp$) is redistributed by pitch angle scattering to 3 degrees of freedom ($3dT$)  the conservation of energy implies

\begin{equation}
3dT=2TB^{-1}dB
\label{eq2}
\end{equation}
for $T=T_\perp=T_\parallel$.
This can be easily integrated to give
\begin{equation}
\frac{T}{B}=\frac{T_0}{B_0}\left(\frac{B_0}{B}\right)^{1/3} \label{eq3}
\end{equation}
which predicts a dip of $T/B$ where $B$ has a maximum.

\begin{figure}
\includegraphics[width=\columnwidth]{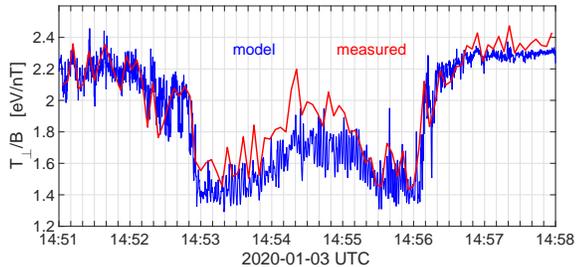}
\caption{Comparison of the measured ratio  $T_{e\perp}/B$ (red) and the modeled dependence expressed by Eq. (\ref{eq3}) (blue) for the second event in Fig. \ref{Fig1}. It shows excellent agreement in the ramp of the shock, where  (\ref{eq3})  is applicable. \label{Fig5}}
\end{figure}

Figure \ref{Fig5}  shows a detailed comparison of Eq. (\ref{eq3}) with the measured ratio for the second  shock structures of Fig. \ref{Fig1}. The model is in excellent agreement with measurements, which supports the validity of the above described type of non-adiabatic heating, henceforth referred to as \emph{quasi-adiabatic heating}. All shock crossing in Fig. \ref{Fig1} show similar signatures of  electron quasi-adiabatic heating with  $\chi_e<$1.  The agreement between the two curves in Fig. \ref{Fig5} indicates an outstanding quality of the particle measurements by the FPI instrument \citep{Pollock:2016}, which is able to reproduce the subtle effects of Eq.~(\ref{eq3}) at sharp gradients of the shock ramps seen in Fig.~\ref{Fig1}. 

It comes as a surprise from this analysis that the strong ECD waves (Fig. \ref{Fig3}) with electric field amplitudes of $\sim$150 mV/m and short wavelengths of $\sim r_e$  do not directly heat electrons. Such expectations have also been expressed by \citet{Mozer:2013}, who noted that the wave potential of the ECD waves significantly exceeds the thermal energy of the electrons, so that some amount of heating would be anticipated.  The electron reluctance to stochastic heating appears to be related to the dependence $\chi_e\propto B^{-2}$, and high values of $B$ in the shock ramp, which keeps $\chi_e<1$.

While the oblique electrostatic electric fields of waves $\sim100-4000$ Hz do not demagnetize the electrons, they appear to participate in the redistribution of the perpendicular kinetic energy of  electrons into the parallel direction. Simulation results with parallel electric fields \citep{Stasiewicz:2020d} indicate that these waves  lead to the isotropization of the electron temperature, as seen in the nine bow shock crossings in Fig.~\ref{Fig1}c.

\section{Simulations of non-adiabatic stochastic heating} \label{sec4}

For a possible stochastic heating of particle species of mass $m$, charge $q$ we have available wave frequencies from dc to 4096 Hz shown in Fig.~\ref{Fig3}, and spatial scales ranging from above $\sim$1000 km for magnetosonic waves to below $r_e\sim 1$ km for ECD waves.
To find out which wave frequencies and spatial scales contribute most to the heating we consider an idealized model in which the magnetic field $B_0$ is in the $z$ direction, and a macroscopic convection electric field $E_{0y}$ drives particles into an electrostatic wave with amplitude $E_{0x}$
propagating  in the $x$-direction. We keep the magnetic field constant to separate purely stochastic heating from the quasi-adiabatic heating discussed above. Thus, in the Doppler frame of the satellite, the drifting plasma is characterized by the convecting electric field and the time-dependent wave electric field. The governing equations are

\begin{align}
   m \frac{dv_x}{dt}&=q E_{0x}\cos(k_x x -\omega t)+q v_y B_0,\label{eq_D1}\\
   m \frac{dv_y}{dt}&=qE_{0y}-q v_x B_0,\\
  \frac{d x}{dt}&=v_x,\qquad
  \frac{d y}{dt}=v_y.\label{eq_D2}
\end{align}
We consider only the 2D plane perpendicular to the magnetic field, since in the absence of parallel electric fields the particles simply stream unperturbed along the magnetic field lines.

The system (\ref{eq_D1})--(\ref{eq_D2}) can be transformed to two different but equivalent forms, one in which we have a time-independent electrostatic wave and a modified convection electric field, and one in which the convective field is eliminated and we have the un-shifted frequency in the plasma frame.
A change of frame into that of the phase velocity of the wave,

\begin{equation}
  v_x=\widetilde{V}_x+\frac{\omega}{k_x}, \qquad
   x=\widetilde{X}+\frac{\omega}{k_x}t
\end{equation}
we obtain the system

\begin{align}
   m \frac{d\widetilde{V}_x}{dt}&=q E_{0x}\cos(k_x \widetilde{X})+q v_y B_0,\\
   m \frac{dv_y}{dt}&=q \widetilde{E}_{0y} -q \widetilde{V}_x B_0,\\
  \frac{d \widetilde{X}}{dt}&=\widetilde{V}_x,\qquad
  \frac{d y}{dt}=v_y,
\end{align}
where the shifted convection electric field is

\begin{equation}
  \widetilde{E}_{0y}= E_{0y}-\frac{\omega}{k_x}B_0.
\end{equation}
In this frame, the electric field is time-independent and governed by the electrostatic potential

\begin{equation}
   \widetilde{\Phi}(\widetilde{X},y)=-\frac{E_{0x}}{k_x}\sin(k_x\widetilde{X})-\widetilde{E}_{0y}y.
\end{equation}
On the other hand, by a change of frame into that of the ${\bf E}\times {\bf B}$-drift velocity,

\begin{equation}
  v_x=V_x+\frac{E_{0y}}{B_0}, \qquad
   x=X+\frac{E_{0y}}{B_0}t
\end{equation}
we obtain instead

\begin{align}
   m \frac{dV_x}{dt}&=q E_{0x}\cos(k_x X -\widetilde{\omega} t)+ q v_y B_0,
   \label{eq_conv1}\\
   m \frac{dv_y}{dt}&= -q V_x B_0,\\
  \frac{d X}{dt}&=V_x,\qquad
  \frac{d y}{dt}=v_y,\label{eq_conv2}
  \end{align}
  where the convecting electric field has been eliminated and absorbed into the frequency in the plasma frame

  \begin{equation}
  \widetilde{\omega}=\omega-k_x \frac{E_{0y}}{B_0}. \label{freq}
\end{equation}
These two different approaches indicate that the model of waves in the plasma frame with the wave frequency (\ref{freq}) that absorbs the convection field is equivalent to the static wave structures superposed with the convection \citep{Stasiewicz:2007}.

Without loss of generality we choose to simulate the system (\ref{eq_conv1})--(\ref{eq_conv2}). A suitable normalization of variables \citep{Karney:1979,Fukuyama:1977,McChesney:1987} with time normalized by $\omega_c^{-1}$, space by $k_x^{-1}$ and velocity by $\omega_c/k_x$ with $\omega_c=q B_0/m$ being the angular cyclotron frequency, gives the system of dimensionless, primed variables,

\begin{align}
   \frac{dv_x'}{dt'}&= \chi\cos( x' - \Omega t')+ v_y',\label{eq_normed1} \\
   \frac{dv_y'}{dt'}&= -v_x',\\
  \frac{d x'}{dt'}&=v_x',\qquad
  \frac{d y'}{dt'}=v_y',\label{eq_normed2}
\end{align}
in which there are only two parameters, the normalized wave frequency in the plasma frame,

\begin{equation}
\Omega =\frac{\widetilde{\omega}}{\omega_c}, \label{eq_om}
\end{equation}
and the stochastic heating parameter, equivalent to (\ref{eq1}),

\begin{equation}
  \chi  = \frac{ m k_x E_0}{q B_0^2}=\frac{k_x}{\omega_c}\frac{E_0}{B_0},
  \label{eq_chi}
\end{equation}
which represents the normalized wave amplitude. An important third parameter is the initial velocity of the particles, since stochastic motion takes place only in restricted regions in phase space \citep{Karney:1979,Fukuyama:1977,McChesney:1987}. For a statistical description of the particles, the initial condition can be described by a Maxwellian distribution function

\begin{equation}
  F=\frac{N}{2\pi v_{T_0}^2}\exp\bigg(-\frac{(v_x^2+v_y^2)}{2 v_{T_0}^2}\bigg),
\end{equation}
where $v_{T_0}=(T_0/m)^{1/2}$ is the initial thermal speed and $T_0$ is the initial temperature. In the normalized variables with $F=N (k_x/\omega_c)^2 F'$ it is written

\begin{equation}
  F'=\frac{1}{2\pi v_{x0}'^2}\exp\bigg(-\frac{(v_x'^2+v_y'^2)}{2 v_{x0}'^2}\bigg)
\end{equation}
where $v_{x0}'=k_x r_c$ is the normalized thermal speed and $r_c=v_T/\omega_c$ is the thermal Larmor radius.
The value of $v_{x0}^{\prime}$ determines the initial temperature in the velocity distribution function, which due to the normalization, is in fact proportional to the ratio of the gyroradius to the wavelength $\lambda=2\pi/k_x$.

\begin{figure*}
\includegraphics[width=18cm]{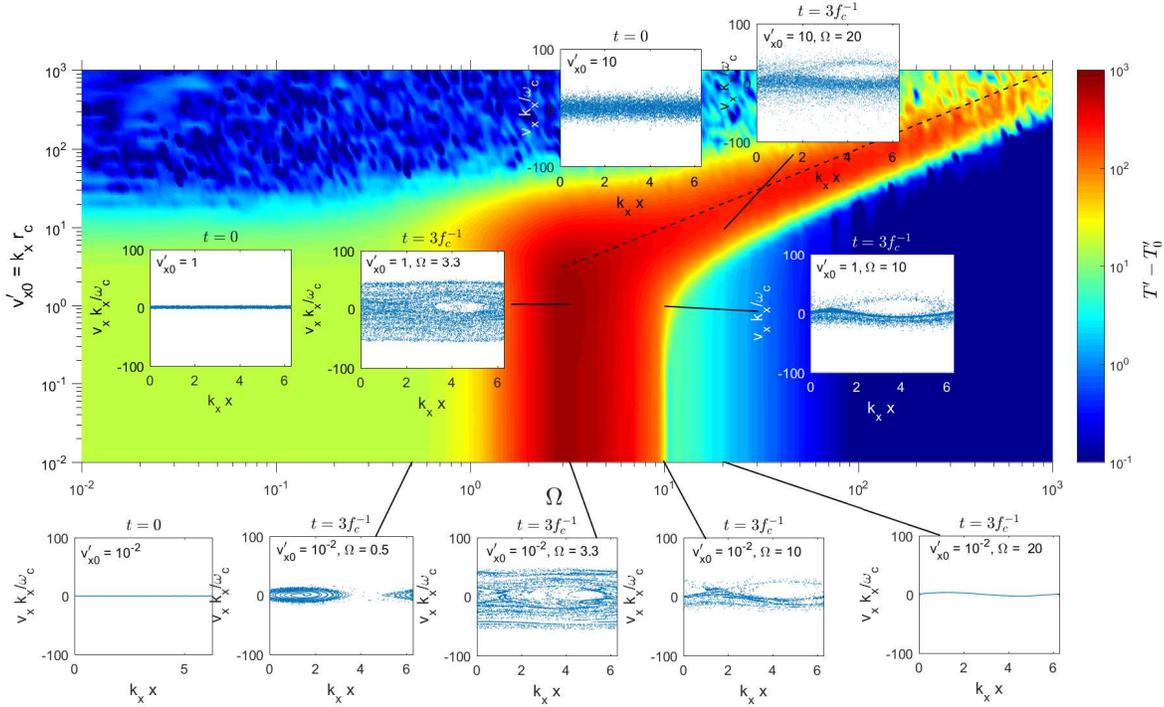}
\caption{A color plot of stochastic heating showing the  difference $T'-T_0'$ between the normalized kinetic temperature $T'=k_x^2 T/m\omega_c^2$ at the end of the simulation and the initial value $T_0'=(v_{x0}')^2=k_x^2 T_0/m\omega_c^2$ after 3 cyclotron periods for charged particles in an electrostatic wave with normalized  amplitude $\chi=60$.
Here $f_c=\omega_c/2\pi$,  $T_0' = v_{x0}'^2$ is the normalized initial temperature and $v_{x0}'=k_x v_{T_0}/\omega_c$ with the thermal speed $v_{T_0}=(T_0/m)^{1/2}$. The insets show distribution functions in ($x$, $v_x$) space at $t=0$ and $3\,f_{c}^{-1}$ for different values of $\Omega$ and $v_{x0}'$. Bulk heating takes place for $\Omega\lesssim 10$, while for $\Omega\gtrsim 10$ there is significant heating only for thermal velocity comparable to the phase velocity, or $v_{x0}'\sim\Omega$ in the normalized variables (dashed line) leading to a distribution function having a high energy tail of particles.  \label{Fig6}}
\end{figure*}

We carry out a set of test particle simulations for  $M=10\,000$ particles, which are Maxwell distributed in velocity and uniformly distributed in space. The system (\ref{eq_normed1})--(\ref{eq_normed2}) is advanced in time using a St{\"o}rmer-Verlet scheme \citep{Press:2007}. The input variables for the simulations are: the normalized wave frequency $\Omega$ in the range $10^{-2}$ to $10^3$, and the initial normalized thermal velocity  $v_{x0}'=k_x r_c$ spanning $10^{-2}$ to $10^3$. The normalized amplitude of the electrostatic wave is set to $\chi=60$, consistent with the observations in Fig.~\ref{Fig4}c. The simulation is run for a relatively short time of 3 cyclotron periods of the particles, motivated by the observations of rapid ion heating within a few cyclotron periods, see Fig. \ref{Fig3}.
The kinetic temperatures resulting from the stochastic heating is calculated as

\begin{equation}
  T=\frac{1}{2M}\sum_{k=1}^M m (v_{x,k}^2+v_{y,k}^2).
\end{equation}

Simulations are carried out for different values of $\Omega$ and $v_{x0}'$ to produce the color plot in Fig.~\ref{Fig6}, which shows the difference $T'-T_0'$ between the normalized kinetic temperature $T'=k_x^2 T/m\omega_c^2$ at the end of the simulation and the initial value $T_0'=(v_{x0}')^2=k_x^2 T_0/m\omega_c^2$.

The most interesting regions are the ones with red color, representing a temperature increase of $\sim (20-1000)\,m\omega_c^2/k_x^2$. The frequency region $1\lesssim\Omega\lesssim10$ represents bulk heating where the cold population is significantly heated to a temperature of $\sim (20-1000)\,m\omega_c^2/k_x^2$. The bulk heating region would expand to higher values of $\Omega$ for larger $\chi$ as well as to lower $\Omega$ for longer times. The inset plots for $\Omega=3.3$ and  initial normalized thermal speeds $v_{x0}'=10^{-2}$ and $1$ show that the particles are bulk heated and spread almost uniformly in velocity space up to a maximum speed $\sim 50\,\omega_c/k_x$, and the distributions achieve a kinetic temperature of $\sim10^3\,m\omega_c^2/k_x^2$ after 3 cyclotron periods. This is relevant for the heating of protons by the low-frequency waves observed in Fig.~\ref{Fig3}. Somewhat similar cases but for $\Omega<1$ and $\chi\sim 1$ leading to rapid heating of ions by drift waves were studied by \citet{McChesney:1987}. For $\Omega=10$ there is also bulk heating  but with a modest increase to about $20\,m\omega_c^2/k_x^2$ after 3 cyclotron periods.

On the other hand, for $\Omega$ significantly larger than $10$ only particles with a high enough initial thermal velocity comparable to the phase velocity, or $v_{x0}'\sim\Omega$ in the normalized variables (dashed line in Fig.~\ref{Fig6})  are further accelerated, leading a warm component with extended energy tails in the distribution function. Such cases of ion heating by lower hybrid waves were discussed by \citet{Karney:1979} and for frequencies near cyclotron harmonics by \citet{Fukuyama:1977}. For $v_{x0}'=10$ and $\Omega=20$ the normalized temperature increases a factor 2 within 3 cyclotron periods, which may be relevant for waves below the lower hybrid frequency seen in Fig.~\ref{Fig3}. Below a threshold initial temperature, the distribution is not affected by the wave, and there is a gap in the heating for low initial temperatures, seen in the lower right blue-colored region of Fig.~\ref{Fig6} including the phase space plots for $v_{x0}'=10^{-2}$ and $\Omega=20$. In this region, the particles oscillate in the wave field without being heated. Finally, for $\Omega\ll 1$ the particles only perform oscillations in an almost time-independent wave electric field, leading to phase-mixing of particles but not to significant stochastic heating.

\section{Discussion}
With the simulation results shown in Section \ref{sec4} we are now in the position to assess which of the broad spectrum of waves in Fig.~\ref{Fig3} are likely to provide stochastic heating of protons at the bow shock.

\begin{figure}
\includegraphics[width=\columnwidth]{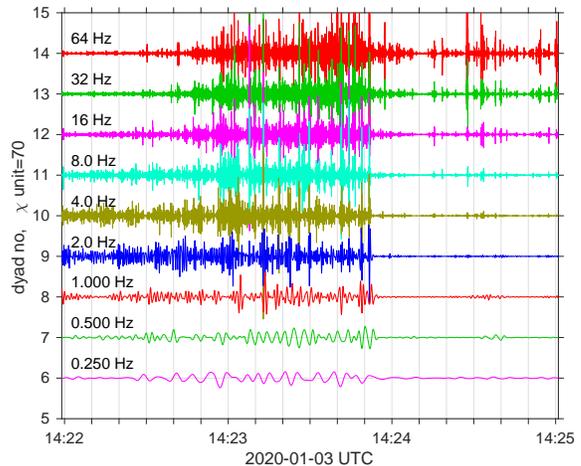}
\caption{Decomposition of $\chi_p$ from Fig. \ref{Fig4}c in the range 0.25-64 Hz shows that heating of protons shown in Fig. \ref{Fig3} can be associated with waves at $f\ge 0.25$ Hz. One plot unit corresponds to $\chi_p= 70$. } \label{Fig7}
\end{figure}

Figure \ref{Fig7} shows the function  $\chi_p$ from Fig. \ref{Fig4}c decomposed into  discrete frequency dyads with orthogonal wavelets \citep{Mallat:1999}.   The signal is divided into discrete frequency layers (dyads) that form $2^{-n}f_N$ hierarchy starting from the Nyquist frequency ($f_N$ is half of the sampling frequency). Orthogonality means that the time integral of the product of any pair of the frequency dyads is zero, and the decomposition is exact, i.e., the sum of all  components gives the original signal. The y-labels are dyad numbers with the unit amplitude corresponding to $\chi_p =70$.
We see that $\chi_p$ in the frequency channels  from 0.25 Hz and above have sufficient amplitude, and correlate well with ion heating seen in Fig. \ref{Fig3} in the time interval 14:23-14:24.
On the other hand, in Fig. \ref{Fig6} we see that bulk proton heating occurs for $f\approx$ (1--10)$f_{cp}$, while the stochastic acceleration of suprathermal particles can be done by waves $f>10f_{cp}$.
Full kinetic simulations of the LHD instability \citep{Daughton:2003} show that the instability develops at longer wavelengths

\begin{equation}
k_\perp (r_e r_p)^{1/2} \approx 1 \label{kr}
\end{equation}
which is equivalent to $k_\perp r_p\approx 10$ in our case,  and has lower frequencies  $f_{cp} < f \lesssim 15f_{cp}$, with significant magnetic component \citep{Gary:1993,Daughton:2003,Huang:2009}. This puts these waves in the bulk heating region of Fig. \ref{Fig6}.  They  could be Doppler upshifted  by 5 Hz, so the possible frequency range for waves that could heat bulk  protons is most likely 0.25-8 Hz in the satellite frame of Figs.~\ref{Fig3} and \ref{Fig7}. Please note that  $k_\perp \equiv k_x$ throughout this paper.

It is the result of the simulations, that the LHD waves at lower frequencies $f_{cp}< f\lesssim 10f_{cp}$, and longer wavelengths $k_\perp r_p \lesssim 30$ are found here to be responsible for the bulk proton heating. Incidentally, they also appear to play a key role in the heating of plasma in the magnetotail and at the magnetopause \citep{Zhou:2014,Graham:2019,Ergun:2019}.

Waves at dyads  16 Hz and above in Figure \ref{Fig7} may correspond  to shorter LHD wavelengths $k_\perp r_e\sim 1$ and frequencies just below $f_{lh}$ \citep{Davidson:1977,Drake:1983}, which are Doppler upshifted to higher frequencies, and also to the modified two-stream instability  which could be triggered by LHDI when the electron drift velocity exceeds the ion thermal velocity \citep{Lashmore:1973,Gary:1987,Umeda:2014}.
This means that they may have $\Omega \sim f_{lh}/f_{cp} \sim 40$ and  $k_xr_c=k_\perp r_p \lesssim 90$ in Fig. \ref{Fig6}. They can be associated with the heating of suprathermal ions in the region $\Omega>10$, around the dashed line $k_x r_c \sim \Omega$ in Fig.~\ref{Fig6}.

\section{Conclusions}
This research has shown that there are two major heating mechanisms in quasi-perpendicular shocks, as implied from the analysis of 9 crossings of the bow shock by the MMS spacecraft.

In this particular event, the electrons do not reach the stochastic heating level with $|\chi_e|<1$ and are  heated by a quasi-adiabatic process related to the compression of the magnetic field at the shock ramp and simultaneous isotropization by LH/ECD waves excited by the density compression. The quasi-adiabatic heating process is supported by the observed isotropic temperature relation $T/B=(T_0/B_0)(B_0/B)^{1/3}$ which predicts a dip in the electron temperature-to-magnetic field ratio when the magnetic field increases.

The ions instead undergo rapid non-adiabatic stochastic heating by  electric field gradients perpendicular to the magnetic field ($\chi_p\approx 60$), and their $T/B$ ratio instead shows an increase in the shock region. Test particle simulations show that efficient stochastic heating within 3 cyclotron periods takes place for a range of parameters in space $(\Omega,\,v_{x0}',\,\chi)$ where $\Omega= \widetilde{\omega}/\omega_c$ is the wave frequency in plasma frame normalized by the cyclotron frequency,  $v_{x0}'=k_x r_c$ is the normalized thermal speed proportional to the ratio between the Larmor radius to the wavelength, and $\chi=(k_x/\omega_c)(E_{0x}/B_0)$ is the stochasticity parameter representing the normalized wave amplitude. We have identified in this space the  range of the ion bulk heating and the range for acceleration of suprathermal ion tails (Figure \ref{Fig6}).

It is found that in the analyzed cases the ion bulk heating is most likely accomplished by  waves in the frequency range 0.25-8 Hz in the spacecraft frame, or (2-10)$f_{cp}$ in the plasma frame, with $k_\perp r_p \lesssim 30$, i.e., with $\lambda>20$ km.  Waves at frequencies larger than 8 Hz in the spacecraft frame ($f>10f_{cp}$ in plasma frame) and with shorter wavelengths can provide acceleration of the tail of the ion distribution function, producing diffuse energetic ion population observed at shocks.

The agreement between the theoretical and numerical results with the MMS observations gives a more complete picture of the heating processes involved in the Earth's quasi-perpendicular bow shock.

The chain of the physical processes described in this paper is initiated by a single event -- namely -- the  compression of the plasma density $N$ and the magnetic field $B$.
 The induced diamagnetic current triggers consecutively three cross-field current driven instabilities: LHD $\rightarrow$ MTS $\rightarrow$ ECD, which produce stochastic bulk heating and acceleration  of ions and electrons, in addition to a common quasi-adiabatic heating of electrons on compressions of $B$.
Thus, the presented model has universal applicability, and the processes described could occur in other types of collisionless shock waves in space, associated with the density compression.  The results may also be applicable  to theories and models of particle heating and acceleration in astrophysical shocks.

\acknowledgments
The authors thank members of the MMS mission for making available the data.
MMS science data is made available through the Science Data Center at
the Laboratory for Atmospheric and Space Physics (LASP) at the
University of Colorado, Boulder: https://lasp.colorado.edu/mms/sdc/public/.
B.E. acknowledges support from the EPSRC (UK), grants EP/R004773/1 and EP/M009386/1.
\software{Data analysis was performed using the IRFU-Matlab analysis package available at https://github.com/irfu/irfu-matlab.}

\bibliographystyle{aasjournal}

\end{document}